\documentstyle[11pt,aaspp4,tighten]{article}
\newcommand{\BV}{Brunt-V\"ais\"al\"a~}
\newcommand{\Pcirc}{P_{\rm circ}}
\newcommand{\curl}{\vec\nabla\times}
\newcommand{\vxi}{\vec\xi}
\newcommand{\dy}{~{\rm d}}
\newcommand{\yr}{~{\rm yr}}

\slugcomment{submitted to ApJ}

\begin{document}


\title{\sc Dynamical Tide in Solar-Type Binaries}

\author{Jeremy Goodman \\
	Eric S. Dickson}

\affil{Princeton University Observatory, 
	Princeton, NJ~08544}

\authoraddr{Jeremy Goodman
Princeton University Observatory, Peyton Hall, Princeton, NJ 08544.
e-mail: jeremy@astro.princeton.edu}

\begin{abstract}
Circularization of late-type main-sequence binaries is usually attributed
to turbulent convection, while that of early-type binaries is explained
by resonant excitation of  g modes.  We show that the latter mechanism
operates in solar-type stars also and is at least as effective as
convection, despite inefficient damping of g modes in the radiative
core.  The maximum period at which this mechanism can circularize
a binary composed of solar-type stars in $10^{10}~\mbox{yr}$ is 
as low as $3$ days, if the modes are damped
by radiative diffusion only and g-mode resonances are fixed;
or as high as $6$ days,
if one allows for evolution of the resonances
and for nonlinear damping near inner turning points.
Even the larger theoretical period falls short of the observed
transition period by a factor two.
\end{abstract}

\keywords{stars: binaries: close, spectroscopic---stars: oscillations}

\section{Introduction}

Coeval spectroscopic binaries show an abrupt transition
between circular orbits at short periods and eccentric orbits at longer
periods.
Furthermore it appears that 
$\Pcirc$ increases with age among main-sequence binaries with roughly
solar-mass components (\cite{MDL92}).

These observations are explained by dissipative tidal interactions.
In eccentric binaries, the tide felt by each star is
time-dependent in a frame rotating with the fluid.
This is also true when the stellar rotation is not synchronous with the
orbit.
Dissipation tends to damp the
variable part of the tide at the expense of orbital or rotational energy,
leading to circularization and synchronization.
Since the tidal force depends strongly on the distance between the
stars---and the tidal dissipation even more strongly---only short-period
binaries synchronize and circularize.

Zahn and collaborators have extensively analyzed two plausible mechanisms
for tidal dissipation (cf. \cite{Z92} for a review and more complete
references).
The ``Theory of the Equilibrium Tide''  assumes
instantaneous hydrostatic  equilibrium in the tidal potential
and attributes dissipation to turbulence in convection
zones (\cite{Z66}, \cite{Z77}).
The theory has been confirmed and roughly calibrated 
by comparison with binaries containing giant stars (\cite{VP96}).
An extension of the theory
explains the small residual eccentricities of binary
pulsars with evolved companions (\cite{P92}).
The predicted transition period for solar-type binaries with ages
$\sim 10^{10}\yr$ is $\Pcirc\approx 2.2\dy$ 
(\cite{GO97}, henceforth Paper I) if one allows for the fact
that the largest eddies turn over more slowly than the binary
orbit and hence do not contribute to tidal dissipation
(\cite{GN77},\cite{GK77}).
Because of the latter effect, the empirical calibrations from
giant stars cannot be applied directly to the main-sequence case,
but even if the inefficiency of slow eddies is ignored,
$\Pcirc$ rises only to $6\dy$ (Paper I).

Notwithstanding theory, observation indicates
that $\Pcirc\gtrsim 11-12~\mbox{d}$ for local disk and halo dwarfs
(\cite{DM91}, \cite{L92}).
Because of the very great sensitivity of tidal circularization
mechanisms to the orbital period, this is a large discrepancy
between observation and theory.

The second of Zahn's mechanisms, the ``Theory of the Dynamical Tide'' 
involves excitation of g modes in the radiative zones of stars
[\cite{Z70}, \cite{Z75}, \cite{Z77}].
Since the orbital period is long compared to the dynamical
time of the star, the g mode has a short radial wavelength, so that
significant coupling between the mode and the tidal potential occurs only
at the boundary between radiative and convective zones where the
radial wavenumber of the g mode vanishes.
This mechanism has been applied to circularization of
early-type binaries and appears to be very satisfactory
(\cite{GMM84}, \cite{CC97}).
These stars have convective cores and radiative envelopes.
It is important for the theory that the thermal timescale 
of the envelope is short, so that
the g modes are efficiently damped.

Having concluded in Paper I that the Theory of the Equilibrium
Tide is inadequate for main-sequence solar-type binaries,
we decided to adapt the Theory of the Dynamical Tide to late-type
systems, in which the cores are radiative and the envelopes convective.
This appears not to have been done before, perhaps because the
damping of g modes in radiative cores is known to be very slow,
which appears at first sight to reduce the efficiency of the mechanism 
sharply (\S3).

Since our methods are similar to Zahn's, we omit many details
and emphasize those aspects that are peculiar to the 
late-type case.
\S 2 treats the excitation of the dynamical tide as if damping
causes the g-mode resonances to overlap.
In fact the resonances do not overlap unless nonlinear damping is efficient,
but the calculation of \S 2 is easy and provides an upper limit to
$\Pcirc$ if the circularization mechanism depends upon the
dynamical tide.
\S 3 considers the (large) corrections that must be made for a finite
star with discrete damped modes.
Finally, \S 4 summarizes our results and discusses the prospects for 
resolving the apparent conflict between theory and observation
concerning $\Pcirc$.

\section{Tidal forcing of g modes}

In a frame corotating with the mean motion of the binary orbit,
the time-dependent quadrupole potential exerted on star $1$ by
the epicyclic motion of star $2$ is, to first order in the
eccentricity ($e$),
\begin{eqnarray}
\delta\Phi_{\rm ext}(t,r,\theta,\phi)
&=& e\, \frac{GM_2}{a^3}r^2 \sqrt{\frac{3\pi}{10}}\\
&\times&\left\{
 Y_{2,2}(\theta,0)[7\cos(\omega t-2\phi)-\cos(\omega t+2\phi)]
-\sqrt{6}Y_{2,0}\cos\omega t\right\},
\label{eq:tpoten}
\end{eqnarray}
where $a$ is the semimajor axis of the binary orbit, 
$\omega\approx \sqrt{G(M_1+M_2)/a^3}$ is the orbital frequency,
and spherical polar coordinates have been chosen with the axis
perpendicular to the orbital plane.

Synchronization is faster than circularization because the
moment of inertia of the individual stars is small
compared to that of the orbit.
Therefore we assume that the stars corotate with the
mean motion of the orbit, but
we neglect coriolis forces so that
we may separate variables in spherical polar coordinates.
The relative error in the dissipation rate caused by this neglect
is expected to be of order unity at most.  Consequently the error in
$\Pcirc$ is expected to be small, since the dissipation
rate varies as a high power of the period.

The fluid displacement in response to the potential (\ref{eq:tpoten})
is conveniently divided into a hydrostatic ``equilibrium tide'' 
$\vec\xi^{\rm eq}$,
which is calculated as though $\omega=0$, and a residual
``dynamical tide'' $\vec\xi^{\rm dyn}$.
In stably-stratified regions,
\begin{equation}
\xi_r^{\rm eq}= -\frac{\delta\Phi}{d\Phi/dr},~~~\mbox{and}~~~
\vec\nabla\cdot\vec\xi^{\rm eq}=0.
\label{eq:eqtide}
\end{equation}
Here $\Phi(r)$ is the potential of the unperturbed spherically-symmetric
star, and $\delta\Phi=\delta\Phi_{\rm ext}+\delta\Phi_{\rm self}$,
where $\delta\Phi_{\rm self}(r,\theta,\phi)$ is the change in the
potential of star 1 due to its distortion.
Equations~(\ref{eq:eqtide}) are consequences of the following facts:
\begin{itemize}
\item[(i)] In hydrostatic
equilibrium, the density and pressure are constant on equipotentials,
and hence the entropy is also constant on equipotentials.

\item[(ii)] In a stratified region, entropy serves as a lagrangian
radial coordinate---that that is, it is conserved by fluid elements
and differs from one mass shell to the next.
\end{itemize}

In unstratified regions (convection zones), the actual
fluid displacement does not follow equations (\ref{eq:eqtide})
in the limit $\omega\to 0$.
Where the entropy and molecular weight are uniform, vorticity is conserved.
For a nonrotating star, therefore, $\curl\vxi=0$, whereas (in general)
$\curl\vxi^{\rm eq}\ne 0$ if $\vxi^{\rm eq}$ obeys
equations (\ref{eq:eqtide}).
We regard equations (\ref{eq:eqtide}) as the definition of 
$\vxi^{\rm eq}$ and require $\vxi^{\rm dyn}\ne 0$ if a convection zone
exists.

For our purposes, we may ignore the selfgravity of both the
equilibrium and the dynamical tides.
The selfgravity of the dynamical tide
is negligible in the convection zone because the mass of the
convection zone of a solar-type star is only a few percent of the
stellar mass; and in the radiative core, the dynamical tide takes
the form of high-order, low-frequency g-modes, whose natural
frequencies are very accurately reproduced in the Cowling approximation
because of their short radial wavelength.
For the equilibrium tide, one
can use the apsidal-motion constant (\cite{CG95})
to estimate that the error in the
tidal dissipation rate caused by neglecting 
$\delta\Phi_{\rm self}^{\rm eq}$
is no more than $10\%$.

One can show that throughout the star, $\xi^{\rm dyn}$ approximately
satisfies the inhomogeneous linear equation
\begin{equation}
\frac{\partial^2}{\partial r^2}(r^2\xi_r^{\rm dyn}) 
+\frac{\partial}{\partial r}\left(\frac{d\ln\rho }{dr}
~r^2\xi_r^{\rm dyn}\right)+
\ell(\ell+1)\left(\frac{N^2}{\omega^2}-1\right)\xi_r^{\rm dyn}
=\ell(\ell+1)\xi^{\rm eq}-\frac{\partial^2}{\partial r^2}(r^2\xi^{\rm eq}).
\label{eq:xirdyn}
\end{equation}
In addition to the approximations discussed earlier (neglect 
of Coriolis forces and $\delta\Phi_{\rm self}$) we have discarded
certain terms that are smaller than those displayed
by factors of $\omega^2/\omega_*^2$, since
the tidal frequencies of interest are small compared with
the the dynamical frequency of the star,
\begin{equation}
\omega_*\equiv\left(\frac{GM_1}{R_1^3}\right)^{1/2}.
\end{equation}
Thus equation (\ref{eq:xirdyn}) does not describe p modes, but only g modes.
The righthand side of eq.~(\ref{eq:xirdyn}) is essentially
the curl of the equilibrium tide and serves as a source
for the dynamical tide in this formulation.

\begin{figure}[h]
\plotone{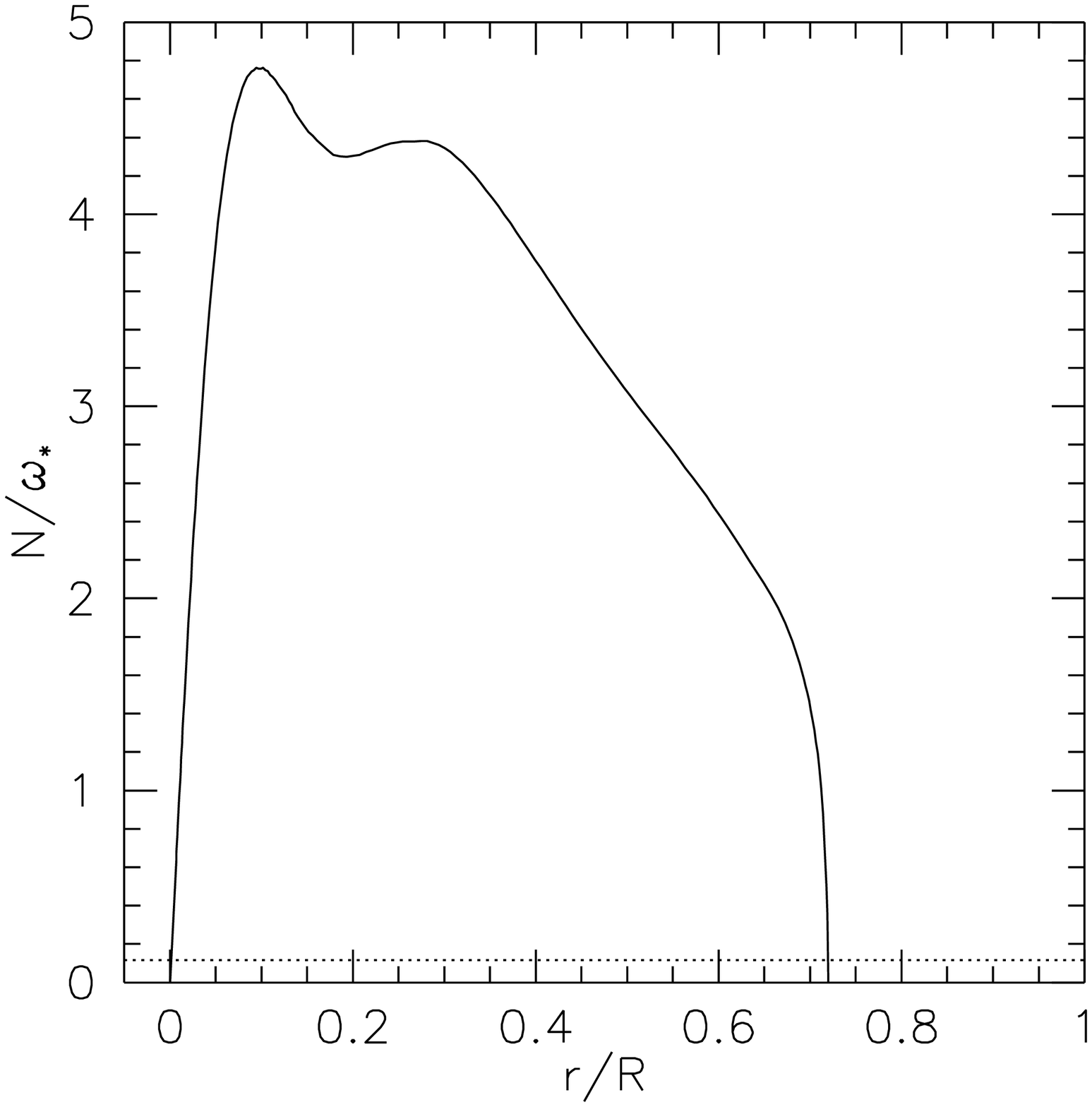}
\caption{\emph{Solid curve}:
\emph{Solid curve}: \BV frequency $N$ normalized to dynamical frequency 
$\omega_*\equiv(GM_\odot/R^3)^{1/2}$ vs. radius, 
based on solar model of BP.  \emph{Dashed line}:
tidal frequency on same scale
for $P=1\dy$, \emph{i.e.} $\omega/\omega_*=0.116$.
} \label{cap:fig1}
\end{figure}

The square of \BV frequency,
\begin{equation}
N^2(r) \equiv -\frac{1}{\rho}\frac{dP}{dr}\frac{d}{dr}
\ln\left(\frac{P^{1/\Gamma_1}}{\rho}\right),
\label{eq:bveqn}
\end{equation}
is positive and $\sim\omega_*^2$ throughout most of the
radiative core.
In the convection zone, we take $N^2$ to vanish,
although properly it has a small negative value there,
and hence $N^2(r)\to 0$ at the boundary between the
core and the envelope, which occurs at $r_c=0.712 R_\odot$ in 
the solar model of \cite{BP95} (henceforth BP).
The run of $N(r)$ derived from their model is shown in
Figure~\ref{cap:fig1}.

Throughout most of the core, the term $(N/\omega)^2$ on the lefthand
side of eq.~(\ref{eq:xirdyn}) is very large, since we are concerned
with tidal frequencies $\omega\ll\omega_*\sim N$.
Consequently,
$\xi_r^{\rm dyn}$ has a wavelength $\ll r$ and is accurately described
by a WKBJ approximation.
The radial wavenumber is
\begin{equation}
k_r\equiv\left[\frac{\ell(\ell+1)}{r^2}
\left(\frac{N^2(r)}{\omega^2}-1\right)\right]^{1/2}.
\label{eq:krdef}
\end{equation}
Near $r=r_c$, the wavenumber vanishes, the mode has a turning point, and
$\xi_r^{\rm dyn}(r)$ can be approximated
by a solution of Airy's equation,
\begin{eqnarray}
\frac{d^2 y}{dx^2} - xy&\approx&0,~\mbox{where}\nonumber\\
y&\equiv& \rho^{1/2} r^2\xi_r^{\rm dyn},\nonumber\\
             x&\equiv& \frac{r-r_c}{\lambda},
\label{eq:Airy}
\end{eqnarray}
and the radial lengthscale
\begin{equation}
\lambda\equiv \left|\frac{\ell(\ell+1)}{\omega^2 r^2}
\frac{dN^2}{dr}\right|_c^{-1/3}\approx 0.025 P_{\rm d}^{-2/3} R_{\sun}.
\label{eq:deltar}
\end{equation}
The numerical value is derived from BP's solar model (Fig.~1), where
$P_{\rm d}$ is the orbital period $2\pi/\omega$ measured in days, and
the subscript $c$ denotes evaluation at $r=r_c$.
On the scale $\lambda$, the righthand side of Eq.~(\ref{eq:xirdyn})
is of order $(\lambda^2/r^2)$ compared to the lefthand side, 
so the Airy equation (\ref{eq:Airy}) has been made homogeneous
for $x<0$ (\emph{i.e.} $r<r_c$).

In the convective envelope ($r_c\le r\le R_1$), 
one must solve the full inhomogenous equation, because
$N^2(r)\approx 0$ so that $\xi_r^{\rm dyn}$ varies on scales $\sim r$.
In fact $\xi_r^{\rm dyn}$ has a constant sign except near
the boundaries:
\begin{eqnarray}
\xi_r&\approx& 0 ~\mbox{at}~r=R_1;\nonumber\\
\xi_r&\approx& 0 ~\mbox{at}~r=r_c.
\label{eq:bcs}
\end{eqnarray}
The first of these conditions reflects the requirement that the surface
be nearly an equipotential, because deviations from an equipotential
have a characteristic frequency $\sim\ell^{1/2}\omega_*\gg\omega$.
The second condition reflects the requirement that the solution
in the convection zone match smoothly at $r_c$ onto the solution in
the radiation zone, for which
\begin{displaymath}
\xi_r^{\rm dyn}\sim\lambda\frac{\partial\xi_r^{\rm dyn}}{\partial r}
\ll r\frac{\partial\xi_r^{\rm dyn}}{\partial r}.
\end{displaymath}

The strongest tidal components are quadrupolar ($\ell=2$), and they
have the form
\begin{displaymath}
\delta\Phi_{\rm ext}=\mbox{Real}\left[
Q r^2 Y_{2,m}(\theta,\phi)e^{-i\omega t}\right],
\end{displaymath}
where $Q$ is a constant with dimensions of $(\mbox{frequency})^2$.
The solution of eqs.~(\ref{eq:xirdyn}) and (\ref{eq:bcs}) in the
convection zone yields
\begin{equation}
\frac{\partial\xi_r^{\rm dyn}}{\partial r}(r_c)\approx
\sigma_c\frac{Q R_1^3}{GM_1},
\label{eq:slope}
\end{equation}
where $\sigma_c$ is a dimensionless constant that depends mainly on
the thickness of the convection zone, and the $(\theta,\phi,t)$
dependence is implicit but understood to be the same as that of
$\delta\Phi_{\rm ext}$.

Equations (\ref{eq:slope}) and (\ref{eq:Airy})
determine the amplitude of the wave
that is launched into the radiative core from $r=r_c$.
This wave propagates inward until it is reflected from
a turning point near $r=0$ and returns to $r=r_c$ with
a phase that depends sensitively on the tidal forcing frequency $\omega$.
If the waves were entirely undamped in the core, their energy would 
increase secularly only when $\omega$ coincided with a global g-mode
eigenfrequency $\omega_{\ell,n}$; otherwise the energy would
saturate after a time $\sim (\Delta\omega)^{-1}$ determined by
the distance $\Delta\omega$ to the nearest resonance.
Thus without damping, there is no secular effect on the binary
eccentricity except at exact resonance.

There is a broad regime of moderate wave damping in which energy is 
transferred 
steadily from the orbit to the star at a rate that is insensitive
to the position of the resonances and independent of the actual
damping rate, $\gamma$.
On the one hand, if $\gamma\ll\omega$, then damping is too small to affect
the tidal excitation of the ingoing wave.
On the other hand, if $\gamma$ exceeds the reciprocal of
the round-trip travel time between the turning points,\
\begin{equation}
t_{\rm group} = 2\int\limits_{r_{\min}}^{\rm r_{\max}}
\frac{dr}{|v_{\rm group}|}\approx
2\int\limits_{0}^{r_c}
\left|\frac{\partial k_r}{\partial\omega}\right|dr,
\label{eq:tgroup}
\end{equation}
then the outgoing wave returning to $r_c$ can be neglected.
The frequency difference between neighboring resonances is
$\Delta\omega=2\pi t_{\rm group}^{-1}$, so the condition 
$\gamma^{-1}\lesssim \pi t_{\rm group}$ ensures that the resonances
are broad enough to overlap.
Hence this regime can be summarized as
\begin{equation}
2\pi t_{\rm group}^{-1}\ll\gamma\ll\omega~~~\mbox{(moderate damping)}.
\label{eq:moderate}
\end{equation}

Under the conditions (\ref{eq:moderate}), the solution 
(\ref{eq:bcs})-(\ref{eq:slope}) in the convection zone must
match onto a solution of the homogeneous Airy equation (\ref{eq:Airy})
corresponding to an inward-propagating wave,
\begin{displaymath}
y(x)\propto \mbox{Bi}(x)-i\mbox{Ai}(x),~~~x\le0.
\end{displaymath}
The mechanical power carried by this wave is
\begin{equation}
\dot E_0= \frac{3^{2/3}}{8\pi}\Gamma^2(1/3)\left[\ell(\ell+1)\right]^{-4/3}
\omega^{11/3}\left[\rho r^5\left|\frac{dN^2}{d\ln r}\right|^{-1/3}
\left|\frac{\partial\xi_r^{\rm dyn}}{\partial r}\right|^2
\right]_{r=r_c}.
\label{eq:eflux}
\end{equation}
With moderate damping, the quotient of $\dot E_0$ by
the energy in the epicyclic motion of the relative orbit,
\begin{equation}
E_{\rm epi}= \frac{1}{2}\frac{M_1M_2}{M_1+M_2} ~e^2\omega^2 a^2 ~+O(e^4),
\label{eq:Eepi}
\end{equation}
yields the rate of decay of the epicyclic energy.
Evaluating $\dot E_0$ from equations (\ref{eq:slope}) and (\ref{eq:eflux}),
and adding the contributions of distinct quadrupolar
spherical harmonics $Y_{2,m}$ in quadrature, we find
\begin{eqnarray}
t_{\rm circ,0}^{-1}&\equiv& -\frac{1}{e}\frac{de}{dt}
=\frac{\dot E_0}{2E_{\rm epi}}\nonumber\\
&\approx& 2.88\sigma_c^2 \left[\rho r^5
\left|\frac{dN^2}{d\ln r}\right|^{-1/3}\right]_{r=r_c}\omega_*^{5/3}
\frac{M_2}{M_1^2 R_1^2}\left(\frac{M_1+M_2}{M_1}\right)^{11/6}
\left(\frac{\omega}{\omega_*}\right)^{7}.
\label{eq:tcdamped}
\end{eqnarray}

We then specialize to the case of solar-type stars and double the
circularization rate to allow for dissipation in both stars.
In BP's solar model, $dN^2/d\ln r\approx -50\omega_*^2$ at $r=r_c$,
and by numerical integration of eqs.~(\ref{eq:xirdyn}) \& (\ref{eq:bcs}),
we find $\sigma_c\approx -0.64$ [cf. eq.~(\ref{eq:slope})].
The result for the circularization time is
\begin{equation}
t_{\rm circ,0}\approx 8.0\times10^3 P_{\rm d}^7\yr.
\label{eq:tcirc0}
\end{equation}
The subscript ``0'' serves as a reminder that this pertains to
the moderate damping regime.

In the simple quasilinear theory of this paper, the eccentricity
decays exponentially but never entirely vanishes.
For ease of reference, we \emph{define} the transition period
$\Pcirc$ by
\begin{equation}
t_{\rm circ}(\Pcirc)\equiv t/3,
\label{eq:tcdef}
\end{equation}
where $t$ is the age of the system, so that
systems with periods $\le \Pcirc$ and initial eccentricities
$e_i\sim 0.3$ have reached $e\lesssim 0.015$.

According to eqs.~(\ref{eq:tcirc0}) \& (\ref{eq:tcdef}), the circularization
period of the oldest solar-type main-sequence stars 
would be $P_{\rm circ,0}\approx 6.4\dy$ in the moderately damped regime.
As we show in the next section,
however, solar-type binaries with comparable periods are \emph{not}
in the moderately damped regime unless nonlinear damping is efficient.

\section{Discrete Modes and Damping}

The derivation of the circularization rate (\ref{eq:tcdamped})
treated the g-mode excited at the edge of the core as though
it were purely an ingoing wave.
In fact, the tidally-forced disturbance is a superposition of
discrete global modes---standing rather than traveling waves---whose
frequencies are quantized:
$\omega_{\rm mode}\in\{\omega_{\ell,n}\}$, where $n\in\{1,2,\ldots\}$
indexes the number of radial nodes.
The damping rate of these modes by radiative diffusion (which is the
dominant linear damping mechanism) is small compared to the difference
$\Delta\omega\equiv\omega_{\ell,n+1}-\omega_{\ell,n}$ between
neighboring eigenfrequencies.
This is quantified by the parameter $\alpha\ll 1$ estimated in
equation (\ref{eq:atten}) below.
The damping is not in the  ``moderate'' regime defined
above---it is much weaker, at least if we consider only
linear damping mechanisms.

So it may fairly be asked why one should bother with the estimates of the
previous section, rather than work with discrete modes from the outset.
Part of the answer is that the tidal power $\dot E_0$ [eq.~(\ref{eq:eflux})]
and the corresponding circularization rate $t_{\rm circ,0}$
[eq.~(\ref{eq:tcdamped})] are equal to the averages of the true
quantities $\dot E(\omega)$ and $t_{\rm circ}(\omega)$ over the discrete
modes, if the average is taken over a range $\gtrsim\Delta\omega$
but $\ll\omega$ itself, and if the average is uniformly
weighted in frequency.  This is an example of a ``sum rule.''  It is
true because $\dot E$ is derived from linear theory and because
if the star is subject to a short-lived tidal disturbance with a lifetime
$\Delta t\ll (\Delta\omega)^{-1}= t_{\rm group}/2\pi$, 
then many discrete modes are excited, but the total dissipation
is given correctly by
the methods of \S 2 since the disturbance ends before the wave
completes a round trip between the turning points.

The second part of the answer is that the estimates of \S 2 are
secure, whereas the consequences of the discreteness of the modes
are complex and somewhat uncertain.
A uniformly-weighted  average of $t_{\rm circ}^{-1}(\omega)$ with respect to 
$\omega$
(yielding $t_{\rm circ,0}^{-1}$) is not appropriate, since the former
is extremely sensitive to the latter---varying by factors 
$\sim\alpha^{-2}\gg 1$ as $\omega$ varies by $\Delta\omega$---and
since $\dot\omega$ is dominated by the tidal interaction itself when
$\omega$ is close to a resonance.
On the one hand,
if the tidal interaction is the only cause of change in the
distance $\min_{n}|\omega-\omega_{\ell,n}|$ from resonance, then one
should average $t_{\rm circ}$, not $t^{-1}_{\rm circ}$, 
uniformly in $\omega$;
this produces the ``harmonic'' mean (\ref{eq:harmonic}), which is
comparable to the maximum value of the instantaneous circularization
time and
is larger than $t_{\rm circ,0}$ by a factor $\alpha^{-1}$.
On the other hand, the distance from resonance is changed by effects
independent of the tide.
One such effect is gravitational radiation; this is negligible
for orbital periods of interest, but not by a very large
factor.
More important is the change in the mode frequencies
$\{\omega_{\ell,n}\}$ as the star evolves and its mean density declines.
Finally, there is good reason to expect that the damping rate should
be enhanced by nonlinear processes, although these are difficult
to estimate.

In this section, we discuss the effects of the last paragraph 
sequentially.
First we consider linear damping by radiative diffusion and derive
the appropriate (harmonic) mean circularization rate if the
only influence on the resonance relations is the tidal interaction
itself.  Then we allow for changes in the resonant frequencies by
stellar evolution.
Lastly, we demonstrate that the dynamical tide is significantly
nonlinear near the center of the star, and explain why this is
likely to cause additional damping.

\subsection{Damping by radiative diffusion}

The density perturbations of a g mode are accompanied by
temperature fluctuations.
Radiative diffusion tends to smooth out the temperature variations
and damp the mode at the local rate
\begin{equation}
\label{eq:localdamp}
\gamma(r)\approx \frac{1}{2}k_r^2\chi(r) \left[1 +~
\frac{\Gamma_1 d\ln\mu/dr}{d\ln(P/\rho^{\Gamma_1})/dr}\right].
\end{equation}
This formula assumes that $\omega\ll N(r)$ so that
the radial wavenumber $k_r\gg r^{-1}$ [cf. eq.~(\ref{eq:krdef})].
The thermal diffusivity is given by
\begin{equation}
\chi\equiv\frac{4acT^3}{3\kappa\rho^2 C_P}=\left(-\frac{dT}{dr}
\frac{dm_r}{dr}\frac{5}{2}\frac{k_B}{\mu m_H}\right)^{-1} L_r,
\label{eq:chi}
\end{equation}
in which all symbols have their conventional meaning.

The stable stratification of the solar core is due mainly
to entropy gradients ($dS/dr>0$) but is reinforced by a
gradient in molecular weight ($d\mu/dr<0$), especially near the center.
The bracketed factor in eq.~(\ref{eq:localdamp})
reflects the fact that temperature perturbations are
associated only with the entropy gradient.  This factor is
less than unity in the core of BP's solar model, but it is
never negative (which would indicate instability to semiconvection).

The global damping rate is the average of the local
rate weighted by the time spent by a wave packet at each radius:
\begin{equation}
\gamma\approx \int\limits_0^{r_c}\frac{\partial k_r}{\partial\omega}
\gamma(r)dr\left/
\int\limits_0^{r_c}\frac{\partial k_r}{\partial\omega}dr\right. .
\label{eq:globaldamp}
\end{equation}
We have approximated the inner and outer turning points by
$0$ and $r_c$, respectively, which is appropriate for
$\omega\ll\omega_*$ and $\ell=2$.

We define the {\it attenuation} ($\alpha$) of the mode by
$\gamma t_{\rm group}$ [cf. eq.~(\ref{eq:tgroup})].
During a roundtrip between the turning points, the
wave amplitude is damped by $\exp(-\alpha)$.
The attenuation determines whether the g-modes are
effectively discrete or continuous:  $\alpha/2\pi$
is the ratio between the half width at half maximum
and the separation of neighboring g-mode resonances at a
common $\ell$.  Thus $\alpha\gtrsim 1$ is the condition for
resonances to overlap.
A numerical calculation in the solar model of BP yields
\begin{equation}
\alpha\approx\left(\frac{P}{11.6\dy}\right)^{4}.
\label{eq:atten}
\end{equation}
The $P^4$ dependence occurs because the local damping rate
$\gamma(r)\propto k_r^2\propto P^2$ [eq~(\ref{eq:krdef})],
and the roundtrip time $t_{\rm group}$ is also $\propto P^2$
[eq.~(\ref{eq:tgroup})].
Clearly $\alpha\ll 1$ and the resonances are well separated at
the circularization period predicted in \S 2 ($6.4\dy$), and
this must be taken into account when one estimates the average
circularization rate.

We can derive an approximate expression for the tidal dissipation
rate from the requirements that 
\begin{itemize}
\item[(i)] $\dot E(\omega)$ should
reduce to $\dot E_0$ when averaged over a range 
large compared to $\Delta\omega\equiv
|\omega_{\ell,n+1}-\omega_{\ell,n}|$, but small
compared to $\omega$ itself.

\item[(ii)] $\dot E(\omega)$ can be expressed as a sum
of Breit-Wigner profiles centered at each resonance.
\end{itemize}
Of course $\dot E_0$ itself depends on $\omega$, but only
as a power-law, whereas $\dot E(\omega)$ varies tremendously
on the scale $\Delta\omega\ll\omega$.
Therefore the average considered in point (i) makes sense
only if $\omega\gg\Delta\omega$, or equivalently, if
the modes with frequecies in the vicinity of $\omega$ are
of large radial order $n$.
In the solar model, $n\approx 100 P_{\rm d}$ for $\ell=2$ modes
(\cite{PB86}), so this condition is well satisfied.

From the two conditions cited, it follows that
\begin{equation}
\dot E(\omega)\approx \dot E_0
\sum\limits_n 
\frac{\gamma\Delta\omega/\pi}{(\omega-\omega_{\ell,n})^2 +\gamma^2}
\end{equation}
where $\gamma$ and $\Delta\omega$ vary slowly with $n$ but can
be treated as constants over ranges $\Delta n\ll n$.
In the same approximation, the infinite sum can be expressed
in closed form as
\begin{eqnarray}
\dot E(\omega)&\approx&
\frac{\dot E_0\sinh\alpha}{2\left[\sin^2 x
+\sinh^2(\alpha/2)\right]}\\
x&\equiv&\left(\frac{\pi}{\Delta\omega}\right)
\min_{n}|\omega-\omega_{\ell,n}|.
\label{eq:edottrue}
\end{eqnarray}

\begin{figure}[h]
\plotone{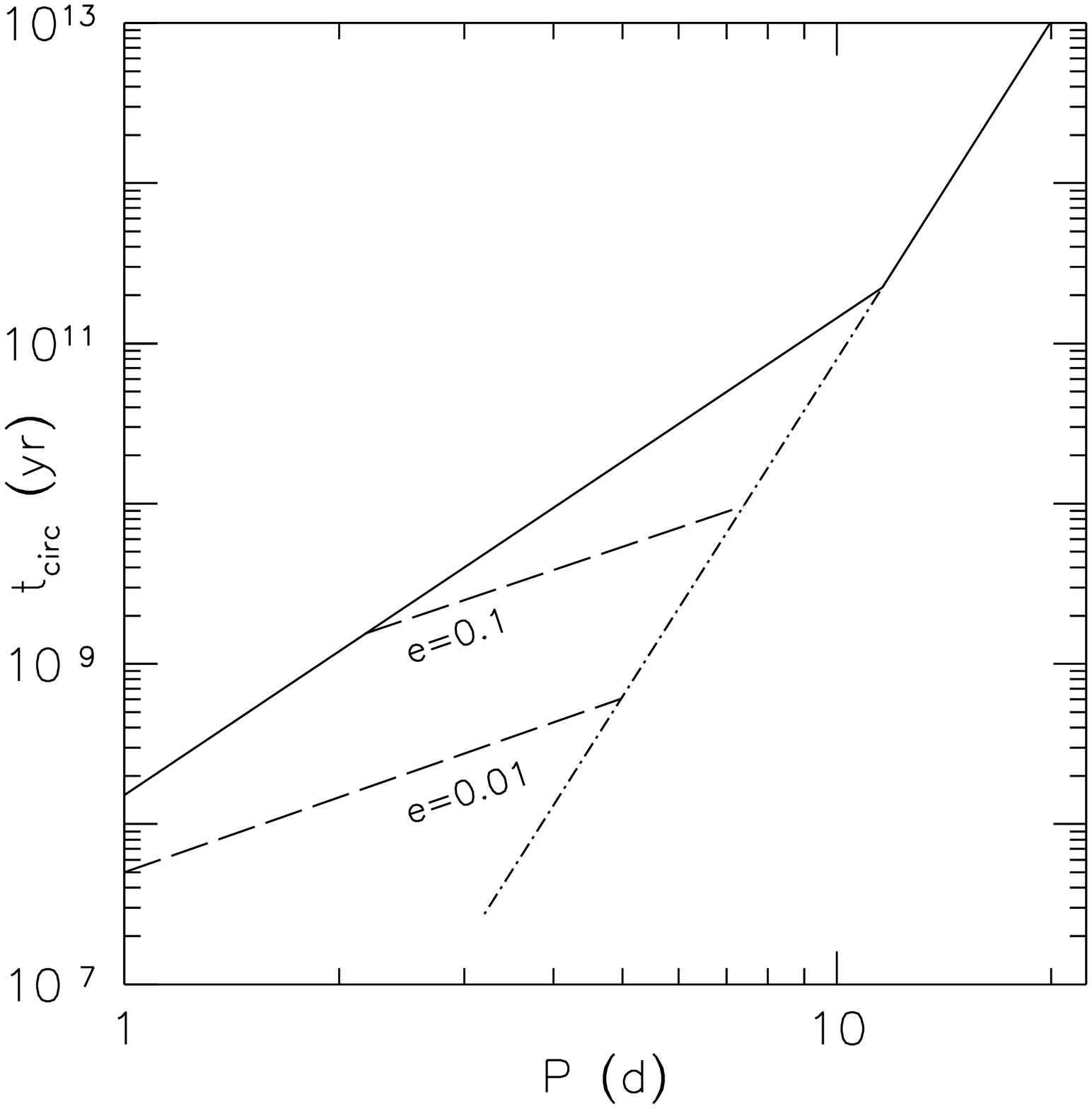}
\caption{
Circularization time, averaged over resonances, versus binary period.
\emph{Broken solid line}:
Fixed resonances and linear radiative damping, based on
eqs.~(\ref{eq:tcirc0}), (\ref{eq:tcirch}), and (\ref{eq:atten}).
\emph{Dot-dashed line}: extension of eq.~(\ref{eq:tcirc0}) to $P<11.6\dy$
given efficient nonlinear damping and/or relatively rapid evolution
of resonant frequencies (\S3.3, \S3.2).
\emph{Dashed lines}: Effect of evolving resonant frequencies for eccentricities
as marked; applicable only when between solid and dot-dashed lines
[eq.~(\ref{eq:tcircev})].
} \label{cap:fig2}
\end{figure}

The tidal frequency is not constant during the history of a close
binary.  The circularization process itself removes the epicyclic energy 
(\ref{eq:Eepi}) from the orbit, causing $\omega$ to increase
at the rate
\begin{equation}
\left(\frac{\dot\omega}{\omega}\right)_{\rm circ}
= \frac{3e^2}{t_{\rm circ}(\omega)}.
\label{eq:omdotcirc}
\end{equation}
Therefore the time spent in the interval
$(\omega,\omega+d\omega)$ of orbital frequency is proportional to
$d\omega/\dot E(\omega)$, and it follows that
the average of $\dot E(\omega)$ with respect to time
is equivalent to its harmonic average with respect to frequency:
\begin{equation}
\langle\dot E\rangle_h = \left.
\int\frac{d\omega}{\dot E(\omega)}\dot E(\omega)\right/
\int\frac{d\omega}{\dot E(\omega)}
=\dot E_0\tanh(\alpha).
\label{eq:harmonic}
\end{equation}
The corresponding circularization time of solar-type binaries is
\begin{equation}
t_{\rm circ,h}\approx \alpha^{-1}t_{\rm circ,0}
\approx 1.5\times 10^8 P_{\rm d}^3\yr~~~(P\ll 11.6\dy),
\label{eq:tcirch}
\end{equation}
and the transition period for systems $10^{10}\yr$ old
is $\Pcirc=2.8\dy$.
This is very much less than the value $6.4\dy$ obtained in \S2
by assuming that the resonances overlap.
The situation is summarized graphically by the solid lines
in Figure~\ref{cap:fig2}.

\subsection{Effects of stellar evolution}

Equations (\ref{eq:edottrue}) and (\ref{eq:harmonic})
imply that the harmonic average of the dissipation rate
is just twice dissipation rate halfway between resonances (\emph{i.e.}
at $x=\pi/2$): 
$\dot E_{\min}\approx (\alpha/2)\dot E_0$.
By contrast, the instantaneous dissipation rate at exact resonance ($x=0$)
is $\dot E_{\max}\approx (2/\alpha)\dot E_0$.
To the extent that $\dot\omega\propto\dot E(\omega)$,
the time required to circularize is clearly dominated by the mininum
value of $|\dot\omega|$ due to the tide, which we denote 
$\dot\omega_{\rm min}$.
Any other process that causes secular changes 
$|\dot\omega|>\dot\omega_{\rm min}$---or even a nonsecular process
that slowly modulates $\omega$ by more than $\Delta\omega/2$----can
substantially influence the mean circularization rate in the regime
$\alpha\ll 1$.

For example, gravitational radiation causes
\begin{equation}
\left(\frac{\dot\omega}{\omega}\right)_{\rm GR}
= \frac{1}{1.7\times 10^{11}\yr}\left(\frac{M_1}{M_\odot}\right)^{5/3}
P_{\rm d}^{-8/3},
\label{eq:omdotgr}
\end{equation}
if the stars have equal masses and $e\ll 1$.
Comparing this with equations (\ref{eq:omdotcirc}) and (\ref{eq:tcirch}),
one sees that gravitational radiation would begin to influence the mean
circularization rate when $e<0.024 P_{\rm d}^{1/6}$.

The ratio of the instantaneous dissipation rate to its time average
depends on $\omega$ through the combination $x\propto\omega-\omega_{\ell,n}$.
Thus changes in the resonant frequencies can be as important 
for $\langle\dot E\rangle$ as changes in $\omega$ (\cite{SP83}).
During the main sequence phase, solar-type stars expand slightly:
for example, \cite{SBK} estimate that the Sun's radius has increased
by about $11.5\%$ since the zero-age main sequence.
If the individual resonant frequencies $\omega_{\ell n}$ scale with
the mean density of the star, then
\begin{equation}
\left(\frac{\dot\omega_{\ell,n}}{\omega_{\ell,n}}\right)_{\rm ev}
\approx -\frac{1}{3\times 10^{10}\yr}.
\label{eq:omdotev}
\end{equation}
Clearly this effect contributes more strongly to $\dot x$ than
gravitational waves, and with the same sign.
It exceeds the minimum value $\dot x_{\rm min}$ due to tidal dissipation 
alone if $e<0.06 P_{\rm d}^{3/2}$, or equivalently if 
$P>2.9 e_{0.3}^{2/3}\dy$, where $e_{0.3}\equiv (e/0.3)\sim 1$ is 
representative of the initial eccentricity (\cite{DM91}).
At substantially smaller eccentricities or longer periods, 
the mean dissipation rate exceeds the harmonic value 
(\ref{eq:harmonic}) by a factor 
$\approx(\dot x_{\rm ev}/4\dot x_{\rm min})^{1/2}$.
This leads to
\begin{equation}
t_{\rm circ,ev}\approx 4.8\times 10^9~e~ P_{\rm d}^{3/2} \yr.
\label{eq:tcircev}
\end{equation}
This value of $t_{\rm circ}$ is valid only if it is less than
the harmonic mean (\ref{eq:tcirch}), else evolutionary effects
on the resonances are negligible, and greater than the
moderately-damped value (\ref{eq:tcirc0}), because as 
$\dot\omega_{\ell,n}\to\infty$, averages over time and frequency
are equivalent, and $\langle\dot E(\omega)\rangle\to\dot E_0$.
Once the eccentricity has been sufficiently reduced so that
the prediction (\ref{eq:tcircev}) is less than (\ref{eq:tcirc0}),
then the latter should be used.

Since the evolutionary changes in $\{\omega_{\ell,n}\}$ begin
to be important only at periods comparable with the $\Pcirc$ obtained
from the harmonic average ($\approx 2.8\dy$), we expect
that $\Pcirc$ is essentially unaffected, except in systems
having a small initial eccentricity.
But the later stages of circularization will proceed on the timescale
(\ref{eq:tcirc0}) rather than (\ref{eq:tcirch}).

\subsection{Nonlinearity}

In linear theory, the turning points occur where $N(r)=\omega$
[cf. eq.~(\ref{eq:krdef})].
Near $r=0$, $N(r)$ is approximately linear in $r$,
and so the inner turning point $r_{\rm min}$ scales as
$P_{\rm d}^{-1}$.
Using the solar model of BP, we estimate that
\begin{equation}
\frac{r_{\rm min}}{R_\odot}\approx 1.25\times 10^{-3} P_{\rm d}^{-1}.
\label{eq:rmin}
\end{equation}
 The smallness of the coefficient reflects the steep gradient of
$N(r)$ near $r=0$ (Fig.~1).

In the WKBJ approximation, the energy flux of the wave is conserved,
so that $\rho r^2 N^2 v_{\rm group}|\bar\xi_r|^2\approx\mbox{constant}$ 
between the turning points.
(Actually a standing wave has zero net flux, since the fluxes of
the ingoing and outgoing wave cancel.  Nevertheless, this argument gives
the correct scaling of the wave amplitude.)
If $\bar\xi_r$ in this formula is the root-mean-square
radial displacement averaged over time and angle at fixed radius,
then the constant is $\dot E_0/4\pi$, where
$\dot E_0$ is given by eq.~(\ref{eq:eflux}) and is related to
$t_{\rm circ,0}$ by equations (\ref{eq:Eepi})-(\ref{eq:tcdamped}).
This allows us to estimate $\bar\xi_r(r)$.
An appropriate dimensionless local measure of nonlinearity is
\begin{equation}
k_r\bar \xi_r \approx 2.8\times 10^{-4} e
\left(\frac{N}{\omega_*}\right)^{1/2} P_{\rm d}^{-11/6}
\left(\frac{r}{R}\right)^{-5/2}.
\end{equation}
At the inner turning point (\ref{eq:rmin}), this becomes
\begin{equation}
k_r\bar\xi_r(r_{\rm min})\approx 1.7\times 10^3 e P_{\rm d}^{1/6}.
\label{eq:nonlin}
\end{equation}
We conclude that the wave is strongly nonlinear near its inner
turning point, unless $e\ll 10^{-3}$.

Sufficiently large-amplitude waves are bound to damp.
Where $k_r\bar\xi_r>1$, for example, the wave inverts 
the stratification during part of its cycle, 
so that denser (lower-entropy) material overlies
less dense (higher-entropy) material.
A Rayleigh-Taylor instability results provided that the
inverted profile persists for more than a growth time, which
translates to $\omega< N$.
The latter condition is not satisfied at $r_{\rm min}$ but it
will be adequately satisfied at radii a few times larger, where
the local nonlinearity is smaller than eq.~(\ref{eq:nonlin}) by a
factor $\approx (r_{\rm min}/r)^{5/2}$.
So to be sure of this instability,
one may take $k_r\bar\xi_r(r_{\rm min})\gtrsim 10$
(say).  If instability does occur, then the energy required to overturn and
mix the material must come at the expense of the wave.
This particular instability may be self-limiting, because mixing will reduce
the molecular-weight gradient, hence move $r_{\rm min}$
outward and reduce $k_r\bar\xi_r(r_{\rm min})$.
Subtler parametric instabilities may arise at weaker nonlinearity 
($k_r\bar\xi_r<1$),
as discussed by \cite{KG}, which are less likely to mix the fluid.

If these or other nonlinear damping mechanisms substantially reduce
the wave reflected from the inner turning point, or even if they
merely perturb its phase in a chaotic manner, then the narrow
resonances pictured earlier will broaden and overlap.
The dynamical tide will then effectively be in the moderate-damping
regime (\ref{eq:moderate}), and in that case the more optimistic
estimates of \S 2 for $t_{\rm circ}$ and $\Pcirc$ should apply.

\section{Discussion}

We have adapted Zahn's Theory of the Dynamical Tide to late-type
stars with radiative cores and convective envelopes.
As in the case of the Theory of the Equilibrium Tide (Paper I),
we have two sets of estimates for the circularization time
($t_{\rm circ}$) and the transition period dividing circular
from noncircular orbits ($\Pcirc$).

Allowing only for the very inefficient linear damping of the tidally
excited g modes by radiative diffusion, we find that the (harmonic mean)
circularization time is given by eq.~(\ref{eq:tcirch}).
Setting $t_{\rm circ}\to 3.3~\mbox{Gyr}$ (one third the main sequence
lifetime of sunlike stars), we obtain a theoretical circularization
period of only 
\begin{equation}
P_{\rm circ,h}\approx 2.8\left(\frac{t}{10^{10}\yr}\right)^{1/3}\dy.
\label{eq:Pcmin}
\end{equation}
Although well short of the observationally inferred values
(cf.  \S 1 and below), this prediction 
is larger than the value $\approx 2.2\dy$ 
predicted by the Theory of the Equilibrium Tide when the latter
is corrected for the slowness of the convective eddies relative
to the tidal period (Paper I).
Adding the two mechanisms raises the prediction only
to $\approx 3.3\dy$ because both mechanisms depend so strongly on $P$.

Because of the combined effects of stellar evolution and nonlinear 
damping, $\Pcirc$ is likely to be larger than eq.~(\ref{eq:Pcmin})
but still smaller than the observed values.
As shown in \S3.3, the linear theory indicates a strongly
nonlinear wave amplitude near the inner turning point as long
as $e\gtrsim 10^{-3} P_{\rm d}^{1/6}$.  If nonlinear losses
substantially reduce the amplitude or randomize the phase
of the wave reflected from the inner turning point, then
we recover the ``moderately-damped'' circularization rate
(\ref{eq:tcirc0}) and the corresponding transition period
\begin{equation}
P_{\rm circ,0}= 6.4\left(\frac{t}{10^{10}\yr}\right)^{1/7}\dy.
\label{eq:Pcirc0}
\end{equation}
On the other hand, once $e\lesssim 0.04$ then stellar evolution
dominates the rate of change of the distance from resonance
even at exact resonance (\S 3.2), 
and one again recovers the larger circularization
rate (\ref{eq:tcirc0}) and transition period (\ref{eq:Pcirc0}).

\cite{MDL92} have concluded that $t_{\rm circ}\propto P^{10/3}$
gives a good fit to the observed scaling of $\Pcirc$ with
binary age.
The exponent of $P$ in the harmonic average (\ref{eq:tcirch}) 
is close to $10/3$.
However, the normalization of $t_{\rm circ}$ that we predict is much
too small to explain the large values of $\Pcirc$ that
\cite{DM91} and \cite{L92} find for old stars ($11-19\dy$).

In summary, theory and observation appear to be in conflict regarding
circularization of late-type main-sequence stars.
The conflict must be resolved either by adjusting the theory, or
by revising the conclusions drawn from the observations:

(i) Perhaps one has overlooked some more effective mechanism of 
circularizing main-sequence binaries at long periods ($\gtrsim 10\dy$).
Tassoul and collaborators  have proposed that rapid circularization
and synchronization of binaries occurs
by a mechanism analogous to Ekman pumping
of fluids in rigid containers (Tassoul 1987, 1995;  
Tassoul \& Tassoul 1996, and references therein).
The reality of this mechanism is controversial on theoretical
grounds (\cite{R92}, \cite{RZ}, \cite{TT97}).
Empirically, Tassoul's mechanism does not
seem to be required to explain circularization of early-type
binaries (\cite{GMM84}, \cite{CC97}) or systems containing
giants (\cite{VP96}), but it contains a free parameter
that Tassoul supposes to vary by orders of magnitude among spectral
types.
Other possibilities include tidally excited parametric instabilities
(\cite{KG}), and tidal excitation of inertial modes, whose existence
depends essentially on coriolis forces (\cite{SP97}); but
neither of these mechanisms has yet been applied to circularization
of late-type binaries.

(ii) Alternatively, perhaps the empirical inference that circularization
occurs on the main sequence is incorrect.
The significance of the trend of $\Pcirc$ with age is not
clear, given the small numbers of binaries with well-determined
masses, ages, and evolutionary states.
Circularization 
is very sensitive to the history of the mean density of the stars,
which can decrease by orders of magnitude in non-main-sequence phases
of evolution (e.g. \cite{VP96}).
There exist old double-lined spectroscopic systems whose members are
almost certainly unevolved with circular
orbits and periods above $10\dy$ (cf. Paper I), but a small number
of such systems might be ascribed to chance.
Furthermore, it is possible that circularization occurs before the main
sequence (\cite{ZB89}).
If the trend of $\Pcirc$ with age is real, it might
result from a difference in mass or metallicity between
the younger disk binaries and older halo systems that correlates with a
difference in pre-main-sequence evolution.
For example, protostellar disks may strongly affect orbital
eccentricity (\cite{ACLP}), and it is conceivable that the persistence
of massive gaseous disks may correlate inversely with the abundance
of dust and other metals in them.

Larger samples of close, well-measured binaries might reduce these 
uncertainties, for example by allowing one to compare systems
of similar mass and metallicity but different age.
Even more helpful
would be large samples of pre-main-sequence binaries, which
it is not unreasonable to expect in the near future (\cite{M94}).

\acknowledgments
We thank Pawan Kumar for helpful discussions.
This work was supported by NASA under grant NAG5-2796.

\end{document}